\documentclass[aps,prb,twocolumn,floatfix,showpacs]{revtex4}
\usepackage{amsmath,amssymb,graphicx,bm}
\begin{document}
\title{Analytic impurity solver with the Kondo strong-coupling
  asymptotics}

\author{V.  Jani\v{s}} \author{P. Augustinsk\'y}

\affiliation{Institute of Physics, Academy of Sciences of the Czech
  Republic, Na Slovance 2, CZ-18221 Praha 8, Czech Republic}
\email{janis@fzu.cz, august@fzu.cz}

\date{\today}


\begin{abstract}
  We present an analytic universal impurity solver for strongly
  correlated electrons. We extend the many-body perturbation expansion
  via suitable two-particle renormalizations from the Fermi-liquid
  regime to the critical region of the metal-insulator transition. The
  reliability of the approximation in the strong-coupling limit is
  demonstrated by reproducing the Kondo scale in the single-impurity
  Anderson model. We disclose the origin of the Kondo resonance in
  terms of Feynman diagrams and find criteria for the existence of the
  proper Kondo asymptotic behavior in approximate theories.
\end{abstract}

\maketitle 

\section{Introduction}
\label{sec:Intro}

One must have at hand a reliable theory capturing the salient features
of the strong-coupling asymptotics to comprehend fully the effects of
electron correlations in metals.  Most theoretical approaches are
based on the well established weak-coupling expansion and we have only
a few opportunities to test its extensions to the strong-coupling
regime. The most reliable check of credibility of a theory in the
strong-coupling regime is to compare its results with one of the
existing exact solutions in this limit.  There are two principal exact
results for strongly correlated electrons. It is the single-impurity
Anderson model (SIAM) and the $1d$ Hubbard model. The ground state of
both models at half filling was constructed with the aid of the Bethe
ansatz. The former is a heavy-fermion liquid in the strong-coupling
limit,\cite{Tsvelik83} while the latter is an electron-hole liquid for
arbitrary interaction strength.\cite{Lieb68}

Recently SIAM has won on importance, since its solutions form a
fundamental building block in the construction of a dynamical
mean-field theory (DMFT) of strong electron correlations via the limit
to infinite-dimensional lattices.\cite{Georges96} Unfortunately, there
are only approximate impurity solvers in the strong-coupling regime.
The only analytic theory, non-crossing approximation, is not a Fermi
liquid and displays low-energy pathological features.\cite{Pruschke89}
Numerical quantum Monte-Carlo simulations\cite{Jarrell92} are
restricted to rather high temperatures.  The most reliable impurity
solver proved to be the numerical renormalization group. It produces
the expected three-peak spectral function with the central Kondo
resonance for intermediate coupling.\cite{Hewson93,Bulla99} Due to
limitations of the numerical procedure it cannot be extended to
arbitrarily large interactions.  Neither of these solutions, however,
is capable to produce a detailed ultimate picture of the the
strong-coupling limit and to decide whether and in which form the
Kondo resonance survives or whether the system undergoes a
metal-insulator transition. This question can only be answered by an
analytically controlled theory.
 
There is presently no global analytic theory capable of tracing down
the genesis of the correlation-induced Kondo resonance The standard
weak-coupling, Fermi-liquid-based perturbation expansion renormalized
with various self-energy insertions becomes unstable for intermediate
and strong electron interaction.\cite{Bickers89} Attempts to go beyond
one-particle renormalizations by introducing explicit vertex
corrections and a two-particle self-consistency either do not lead to
analytically solvable equations or have not yet produced the desired
Kondo asymptotics in SIAM.\cite{Weiner71,Bickers91a}

The aim of this paper is to construct an analytically controllable
approximation that would reliably interpolate between the weak- and
strong-coupling regimes in models with a screened electron repulsion.
To this purpose we use an appropriately renormalized diagrammatic
many-body expansion. We demonstrate that it is not the one-particle
self-consistency but rather a two-particle one that must be introduced
in the critical region of the metal-insulator transition. We further
show that only a symmetric inclusion of electron-electron and
electron-hole scatterings leads to the proper critical behavior. The
two-particle self-consistency will be achieved within the parquet
approximation in which we separate singular and regular functions. Not
to lose analytic control only the potentially divergent functions with
long-range fluctuations are kept dynamical while the regular ones with
short-range fluctuations are replaced by constants. Such a
simplification does not affect the universal features of the critical
behavior. We explicitly construct the approximation for SIAM so that
its reliability becomes transparent by qualitatively correctly
reproducing the exact weak-coupling (Fermi-liquid) and strong-coupling
(Kondo) regimes. We identify with our construction the minimal
necessary conditions on approximate schemes to cover the Kondo
strong-coupling behavior.

\section{Model and  basic equations}\label{sec:Basic}

We formulate the construction of an approximate theory interpolating
between the weak and strong coupling regimes independently of the
underlying model via renormalized Feynman diagrams and equations of
motion.  To demonstrate reliability of our approach we, however, use
in our explicit calculations the single impurity Anderson model the
Hamiltonian of which reads
\begin{multline}\label{eq:H-SIAM}
  \widehat{H} = \sum_{{\bf k}\sigma} \epsilon({\bf k})
  c^{\dagger}_{{\bf k}\sigma} c^{\phantom{\dagger}}_{{\bf k}\sigma} +
  E_d\sum_\sigma d^{\dagger}_\sigma d^{\phantom{dagger} }_\sigma \\ +
  \sum_{{\bf k}\sigma}\left(V^{\phantom{*}}_{{\bf
        k}}d^{\dagger}_\sigma c^{\phantom{\dagger}}_{{\bf k}\sigma} +
    V^*_{{\bf k}} c^{\dagger}_{{\bf k}\sigma}
    d^{\phantom{\dagger}}_\sigma\right) +
  U\widehat{n}^d_\uparrow\widehat{n}^d_\downarrow\ .
\end{multline}
We denoted $\widehat{n}^d_\sigma = d^{\dagger}_\sigma d_\sigma$.  When
calculating the grand potential and thermodynamic properties of this
impurity model we can explicitly integrate over the degrees of freedom
of the delocalized electrons. To this purpose we standardly replace
the local part of the propagator of the mobile electrons by a constant
$\Delta(\epsilon) = \pi \sum_{\bf k} |V_{\bf k}|^2 \delta(\epsilon -
\epsilon({\bf k})) \doteq \Delta $ the value of which we set as the
energy unit. For simplicity we assume half-filled case $\mu = E_d +
U/2$. The impurity grand partition function can then be represented
via a local Grassmann functional integral
\begin{multline}\label{eq:SIAM-GPF}
  \mathcal{Z} = \int \mathcal{D}\psi \mathcal{D}\psi^*
  \exp\left\{\sum_n \psi^*_n (i\omega_n + \bar{\mu} +
    i\text{sign}(\omega_n) \Delta ) \psi_n \right.  \\ \left. - U
    \int_0^\beta d\tau\ \widehat{n}^d_\uparrow(\tau)
    \widehat{n}^d_\downarrow(\tau) \right\}\ .
\end{multline}
For this integral we can straightforwardly build up the standard
weak-coupling (diagrammatic) perturbation expansion in powers of the
interaction strength $U$ (diagrammatically represented as a vertex)
and the bare propagator, being in the impurity case $ G_0(x + i y) =
1/(x + i\ \text{sign}(y)(\Delta + |y|)) $ (diagrammatically
represented as an oriented line).
 
\begin{figure}
  \includegraphics[height=2.3cm]{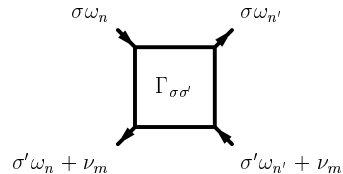}
\caption{\label{fig:2P-generic}Diagrammatic assigning of dynamical
  variables, frequencies and spins, to the vertex
  $\Gamma_{\sigma\sigma'}(i\omega_n,i\omega_{n'};i\nu_m)$.}
\end{figure}

We know that in the particular case of SIAM the non-renormalized
weak-coupling expansion is preferred and converges for arbitrary
interaction strength.\cite{Zlatic83} To enable a treatment of phase
transitions and singularities in more general extended lattice models
we have, however, to renormalize the bare expansion. The fundamental
quantity for renormalization in the many-body perturbation expansion
is the self-energy. In our approach we do not use the self-energy as
the approximation-generating quantity determined directly from sums of
selected (renormalized) diagrams. Instead, we represent it by means of
the Schwinger-Dyson equation and the two-particle vertex $\Gamma$ as
follows
\begin{multline}
\label{eq:sigma-2P} \Sigma_\sigma(i\omega_n)=\frac{U}{\beta }\sum_{n'}
G_{-\sigma}(i\omega_{n'}) \bigg[ 1 -\frac{1} {\beta}\sum_{m}
G_\sigma(i\omega_{n + m}) \\ \times G_{-\sigma}(i\omega_{n'+ m})
\Gamma_{\sigma-\sigma}(i\omega_{n + m};i\omega_n,i\nu_{ n' - n}
)\bigg]
\end{multline}
where $\omega_n = (2n + 1)\pi T$ and $\nu_m = 2\pi m T$ are fermionic
and bosonic Matsubara frequencies in units $k_B=1$, respectively. We
assigned dynamical variables to the two-particle vertex as shown in
Fig.~\ref{fig:2P-generic}.

We further introduce two-particle irreducible vertices, that is,
vertices that cannot be diagrammatically disconnected by cutting
specific pairs of one-particle lines (propagators).  The choice of the
two-particle irreducibility is ambiguous.\cite{Janis99b} Here we
choose only the singlet electron-hole and electron-electron scattering
channels.  We denote the respective irreducible vertices
$\Lambda^{eh}$ and $\Lambda^{ee}$. With their aid we represent the
full two-particle vertex via two nonequivalent Bethe-Salpeter (BS)
equations.  The Bethe-Salpeter equation in the $eh$ channel reads
\begin{subequations}\label{eq:BS}
\begin{multline}\label{eq:BS-eh}
  \Gamma_{\uparrow\downarrow}(i\omega_n,i\omega_{n'},i\nu_m) =
  \Lambda^{eh}_{\uparrow\downarrow}(i\omega_n, i\omega_{n'}, i\nu_m)\\
  - \frac 1\beta
  \sum_{n''}\Lambda^{eh}_{\uparrow\downarrow}(i\omega_n, i\omega_{n'};
  i\nu_m) G_\uparrow(i\omega_{n''})
  G_\downarrow(i\omega_{n'' + m})\\
  \times \Gamma_{\uparrow\downarrow}(i\omega_{n"}, i\omega_{n'};
  i\nu_m)\ .
\end{multline}
In the $ee$ channel we obtain
\begin{multline}\label{eq:BS-ee}
  \Gamma_{\uparrow\downarrow}(i\omega_n, i\omega_{n'}; i\nu_m) =
  \Lambda^{ee}_{\uparrow\downarrow}(i\omega_n, i\omega_{n'}; i\nu_m)
  \\ - \frac 1\beta \sum_{n''}
  \Lambda^{ee}_{\uparrow\downarrow}(i\omega_n,
  i\omega_{n''}; i\nu_{m + n'  -  n''}) G_\uparrow(i\omega_{n''})\\
  \times G_\downarrow(i\omega_{n + n' + m - n''})
  \Gamma_{\uparrow\downarrow}(i\omega_{n''}, i\omega_{n'}; i\nu_{m + n
    - n''})\ .
\end{multline}\end{subequations}

Equations~\eqref{eq:sigma-2P} and~\eqref{eq:BS} reduce the solution of
the investigated model to the determination of the irreducible
vertices $\Lambda^{eh}$ and $\Lambda^{ee}$. The principal idea of the
parquet approach is to use the topological nonequivalence of different
Bethe-Salpeter representations of the full two-particle vertex via the
irreducible ones. Due to the nonequivalence of different types of
two-particle scatterings the reducible vertex in one channel is
irreducible in the other one.\cite{Janis99b} If the vertex irreducible
in both channels is the bare interaction $U$ (parquet approximation)
we obtain the basic parquet equation in our two-channel scheme
\begin{equation}\label{eq:Parquet}
  \Gamma_{\uparrow\downarrow} = \Lambda^{eh}_{\uparrow\downarrow} +
\Lambda^{ee}_{\uparrow\downarrow}  - U\ .
\end{equation}
We use this representation in Eqs.~\eqref{eq:BS} to exclude the full
vertex $\Gamma$ from the BS equations that then become coupled
convolutive nonlinear integral equations for the irreducible vertices
$\Lambda^{eh}$ and $\Lambda^{ee}$.  These equations cannot be solved
analytically and their numerical solution is available only far from
the Kondo regime.\cite{Bickers92} To gain an analytic assessment of
the low-temperature, strong-coupling behavior of the parquet
equations~\eqref{eq:BS} using Eq.~\eqref{eq:Parquet} we have to resort
to simplifications.

\section{Kondo asymptotics -- vertex functions} \label{sec:Kondo}

\subsection{Simplified parquet equations}\label{sec:Simplified}

Equations~\eqref{eq:BS} can be analytically solved (in an approximate
way) individually for the full vertex $\Gamma$ by neglecting
Eq.~\eqref{eq:Parquet} and using the irreducible vertices as input.
Approximations of this type with renormalized one-particle propagators
are now called fluctuation-exchange (FLEX).  In the simplest case the
irreducible vertices are replaced by the bare interaction. We find
that vertex $\Gamma$ tends to develop a pole in the BS equation with
multiple electron-hole scatterings, Eq.~\eqref{eq:BS-eh}, caused by
the imminent metal-insulator transition at zero temperature. While the
two-particle vertex $\Gamma$ remains finite (bounded) when calculated
from Eq.~\eqref{eq:BS-ee}. We use the fact that only
equation~\eqref{eq:BS-eh} develops a pole in simplifying the parquet
approximation in the critical region of the metal-insulator transition
to an analytically controllable theory.  Since the output of the BS
equation from one channel enters the kernel of the other one, only
$\Lambda^{ee}$ becomes singular in the critical region of the
metal-insulator transition.  The irreducible vertex $\Lambda^{eh}$
remains bounded everywhere in the parquet approximation.

We restrict our interest to the leading critical behavior of the
solution of the parquet equations in the asymptotic region of the
metal-insulator transition. To this purpose we use the reasoning of
the renormalization group and neglect the short-range fluctuations. We
take into account explicitly only long-range (in time) fluctuations
shaping the singular behavior. In this way we do not influence the
universal part of the critical asymptotics. We hence keep only
singular functions dynamical (dependent on the relevant frequency
only). Regular functions, where only short-range fluctuations
contribute, can be replaced by constants, suitably chosen averaged
values. We must further guarantee that only the leading low-frequency
behavior of singular functions determines the critical behavior in the
Bethe-Salpeter equations.

As a first step we approximate $\Lambda^{eh}$ with a static effective
interaction $\Lambda^{eh} = \overline{U}$. Inserting this ansatz into
Eq.~\eqref{eq:BS-eh} we obtain a FLEX-type equation for the
irreducible vertex from the electron-electron channel
\begin{subequations}\label{eq:Lambda-ee}
\begin{equation}\label{eq:Lambda-ee-full}
\Lambda^{ee}_{\uparrow\downarrow}(i\omega_n,i\omega_{n'}; i\nu_m)  = U -\
  \frac{\overline{U}^2\chi_{eh} (i\nu_m)}{1 + \overline{U}\chi_{eh} (i\nu_m)}\
\end{equation}
where we denoted the electron-hole bubble $\chi_{eh} (i\nu_{m}) =
\beta^{-1} \sum_n G_\uparrow(i\omega_n) G_\downarrow(i\omega_{n +
  m})$.  Since the static approximation to the electron-hole
irreducible vertex $\Lambda^{eh} = \overline{U}$ causes a deviation
from the exact high-frequency limit $\Lambda^{eh} \to U$,
approximation~\eqref{eq:Lambda-ee-full} holds in the low-frequency
limit.  We hence have to consider in our simplification only the
leading low-frequency contribution from the right-hand side of
Eq.~\eqref{eq:Lambda-ee-full}. It reads
\begin{align}\label{eq:Lambda-ee-simplified}
  \Lambda^{sing}_{\uparrow\downarrow}(i\nu_m)& = -\ 
  \frac{\overline{U}^2\chi_{eh} (i\nu_m)}{1 + \overline{U}\chi_{eh}
    (i\nu_m)}\ .
\end{align}
\end{subequations}

We now use Eqs.~\eqref{eq:BS-ee} and \eqref{eq:Parquet}, and
$\Lambda^{sing}$ from Eq.~\eqref{eq:Lambda-ee-simplified} to determine
the effective interaction $\overline{U}$. The right-hand side. of
Eq.~\eqref{eq:BS-ee} remains frequency-dependent even with our ansatz.
Since we neglect all finite (short-time) fluctuations in regular
functions, we have to replace the vertex $\Gamma$ resulting from
Eq.~\eqref{eq:BS-ee} with a constant.  This replacement is not
uniquely defined but the ambiguity has no impact on the qualitative
(universal) critical behavior of the solution. We found that the most
appropriate way in replacing the full vertex with a constant is to
multiply both sides of Eq.~\eqref{eq:BS-ee} with pairs of one-electron
propagators $G(i\omega_n)G(i\omega_{m-n})$ from the left and by
$G(i\omega_{n'})G(i\omega_{m- n'})$ from the right. We then average
over the fermionic Matsubara frequencies with a constraint $m = 0$,
conserved during the electron-electron multiple scatterings. Doing so
we obtain an explicit equation for the effective interaction, the
static part of the electron-hole irreducible vertex,
\begin{equation}\label{eq:U-renormalized}
\overline{U}\chi_{ee}(0)  =  U\chi_{ee} -\frac{\left\langle
G_\uparrow G_\downarrow L_{\uparrow\downarrow}^2\right\rangle}
{\chi_{ee} +\left\langle G_\uparrow G_\downarrow
L_{\uparrow\downarrow}\right\rangle}\ .
\end{equation}
We denoted $\chi_{ee} = \beta^{-1} \sum_n G_\uparrow(i\omega_n)
G_\downarrow(i\omega_{ - n})$ the static part of the electron-electron
bubble and
\begin{subequations}\label{eq:Convolutions-def}
\begin{equation}\label{eq:Convolutions-two}
L_{\uparrow\downarrow}(i\omega_n) = \frac 1{\beta}
\sum_{n'.}G_\uparrow(i\omega_{n'}) G_\downarrow(i\omega_{-n'})
\Lambda^{sing}_{\uparrow\downarrow}(i\nu_{-n -n'}) \ ,
\end{equation}
\begin{equation} \label{eq:Convolutions-three}
\left\langle G_\uparrow G_\downarrow  X\right\rangle = \frac 1{\beta}
\sum_{n.}G_\uparrow(i\omega_{n}) G_\downarrow(i\omega_{-n})
X(i\omega_{n})\ .
\end{equation} \end{subequations}
Equations~\eqref {eq:Lambda-ee-simplified} -
\eqref{eq:Convolutions-def} form a closed set of relations determining
the static effective interaction $\overline{U}$ and the dynamical
vertex $\Lambda^{sing}(i\nu_m)$ as functionals of the one-particle
propagators $G_\sigma$ and the bare interaction $U$. Together with
Eqs.~\eqref{eq:Parquet} and~\eqref{eq:sigma-2P} we have an analytic
approximation in closed form.  The one-electron propagators may be
either bare or renormalized with the self-energy, when one-particle
self-consistency is used. The latter is not mandatory and we show that
it worsens the approximation with the Hartree one-electron
propagators, in particular in the intermediate and high-frequency
sectors.

The critical region of the metal-insulator transition in SIAM
corresponds to the strong-coupling regime, $U/\Delta \to\infty$.  It
is reached when the denominator on the r.h.s. of
Eq.~\eqref{eq:Lambda-ee-simplified} approaches zero.  At zero
temperature, half filling, and in the rotationally invariant case,
$G_\uparrow = G_\downarrow$ we define a dimensionless Kondo scale $a =
1 + \overline{U}\chi_{eh}(0) = 1 - \overline{U}\int_{-\infty}^0
d\omega \Im[G_+(\omega)^2]/\pi \to 0$.  We used an abbreviation
$G_\pm(\omega) \equiv G(\omega \pm i0^+)$.  The denominator of the
r.h.s. of Eq.~\eqref{eq:Lambda-ee-simplified} can then be expanded
only to the lowest (linear) order in frequency if we are interested in
the leading singular behavior. We obtain $1 +
\overline{U}\chi_{eh}(\omega + i0^+) = a -i\pi
\overline{U}\rho_0^2\omega$. We denoted the density of one-particle
states at the Fermi energy $\rho_0 = -\pi^{-1}\Im G_+(0)$. It does not
depend at zero temperature and half filling on the interaction
strength. This singular low-frequency asymptotics of $\Lambda^{sing}$
allows us to evaluate the leading contributions to the integrals in
Eqs.~\eqref{eq:Convolutions-def}.

The leading contributions to the integrals with vertex
$\Lambda^{sing}$ in the asymptotic limit $a \to 0$ read
 \begin{subequations}\label{eq:Explicit-convolutions}
 \begin{equation}\label{eq:L_ee}
L_{\uparrow\downarrow}(z)  \doteq \frac {G(z) G(-z)}{\pi^2 \rho_0^2}
|\ln a |\ ,
\end{equation}
\begin{multline}\label{eq:Explicit-Lambda-U}
  \left\langle G_\uparrow G_\downarrow L_{\uparrow\downarrow}\right\rangle \\
  \doteq - \left(\frac{ |\ln a|}{\pi^2\rho_0^2} \right)
  \int_{-\infty}^0 \frac {d\omega}{\pi} \Im\left[ G_+(\omega)^2
    G_-(-\omega)^2\right]\ ,
\end{multline}
and
\begin{multline}\label{eq:Explicit-Lambda-Lambda}
  \left\langle G_\uparrow G_\downarrow
    L_{\uparrow\downarrow}^2\right\rangle \\ \doteq -\left(\frac{ |\ln
      a|}{\pi^2\rho_0^2} \right)^2 \int_{-\infty}^0 \frac
  {d\omega}{\pi} \Im\left[ G_+(\omega)^3 G_-(-\omega)^3\right]\ .
\end{multline} \end{subequations}
The effective interaction has an asymptotic solution for $a\to 0$
\begin{equation}\label{eq:Ubar-solution}
\overline{U} = \frac 1{\chi_{ee}} \left[ U\chi_{ee}   - \frac{|\ln
a|}{\pi^2\rho_0^2}\ \frac{\left\langle
(GG)_{ee}^3\right\rangle}{\left\langle (GG)_{ee}^2\right\rangle  }\right]
\end{equation}
where we denoted $\left\langle (GG)_{ee}^n\right\rangle = -
\pi^{-1}\int_{-\infty}^0 d\omega \Im\left[ G_+(\omega)^n
  G_-(-\omega)^n\right]$.

The dimensionless Kondo scale measuring the distance from the
metal-insulator transition is determined from an equation for the
critical point of the metal-insulator transition. It reads $1 +
\overline{U}\chi_{eh}(0) =0$. At half filling $\chi_{eh}(0) =
-\chi_{ee}$ and we obtain a leading-order asymptotic solution for the
distance to the critical point
\begin{equation}\label{eq:a-solution-general}
  a = \exp\left\{ - \pi^2\rho_0^2 \ \frac{ \left\langle
        (GG)_{ee}^2\right\rangle }{\left\langle
        (GG)_{ee}^3\right\rangle} \left[ U\chi_{ee} -
      1\right]\right\}\ .
\end{equation}
Solution~\eqref{eq:a-solution-general} holds at zero temperature and
half filling when the argument in the exponential tends to infinity
and $U\chi_{ee} > 1$. The latter condition can be viewed as a
characteristic of the strong-coupling regime, since the FLEX,
weak-coupling solutions are distinguished by the opposite inequality,
$U\chi_{ee} < 1$.

\subsection{Hartree propagators}\label{sec:Hartree}

To manifest analytically that Eq.~\eqref{eq:a-solution-general}
reproduces correctly the Kondo asymptotics we use the bare (Hartree)
one-particle propagators in Eq.~\eqref{eq:a-solution-general}. In this
case the integrals with powers of the one-particle propagators can be
explicitly evaluated, $\left\langle (GG)_{ee}^n\right\rangle =
\pi^{2(n-1)}\rho_0^{(2n-1)}/(2n-1)$. Using this result we obtain an
explicit representation for the Kondo scale
 \begin{equation}\label{eq:a-solution-nonself}
 a = \exp\left\{ - \frac 53 \left[U\rho_0 - 1
\right]\right\}\ .
 \end{equation}
 The exact result is $a = \exp\{-\pi^2 U\rho_0/8\}$. Determination of
 the exact prefactor at the linear dependence of the exponent of the
 Kondo scale on the interaction strength is beyond the reach of the
 present approximation. The prefactor depends on the way in which we
 replace regular frequency-dependent functions with constants. The
 universal feature of the Kondo scale is \textit{linearity} in the
 interaction strength of the exponent. This linearity is the major
 achievement of our construction. The other Fermi-liquid-based
 approximations fail in the strong-coupling regime of SIAM. The
 single-channel FLEX-type approximations either do not see the
 metal-insulator transition at all ($ee$ channel) or predict a
 quadratic exponent $a \sim \exp\{-U^2\rho_0^2\}$ ($eh$ channels).
 \cite{Hammann69}

 \section{Kondo asymptotics -- one-particle functions}\label{sec:1P-functions}
 
 The core of the parquet approach is a self-consistency determining
 the two-particle irreducible vertices from the completely irreducible
 vertex (bare interaction in the parquet approximation) and the
 one-particle propagators. Both, the completely irreducible vertex and
 the one-particle propagators serve as input to the parquet equations.
 To complete the parquet approximation and make it conserving and
 thermodynamically consistent we have to determine the physical
 one-particle functions, that is, the physical one-particle propagator
 and the self-energy. All physical quantities in conserving theories
 are then generated from the full one-particle propagator or the
 self-energy and their dependence on appropriate external
 sources.\cite{Baym62} The physical one-particle propagator need not
 be identical with the "auxiliary" one used in the parquet equations
 for the two-particle vertices.

\subsection{Self-energy in the parquet approach}\label{sec:Self-energy}

The fundamental quantity for any thermodynamically consistent and
conserving approximation is the self-energy. Knowing the full
two-particle vertex from the parquet equations we determine the
self-energy from the Schwinger-Dyson equation~\eqref{eq:sigma-2P}. It
reduces in the critical region of the metal-insulator transition to a
simple FLEX-type expression
\begin{equation}
\label{eq:Parquet-Sigma} \Sigma_\sigma(i\omega_n)=\frac{U}{\beta }\sum_{n'}
\frac {G_{-\sigma}(i\omega_{n'}) }{1 + \overline{U}\chi_{eh}(i\nu_{n -
n'})}\ .
\end{equation}
We can analytically continue the sum over Matsubara frequencies to an
integral with Fermi function. At zero temperature the real part of the
self-energy reads
 \begin{subequations}\label{eq:Self-energy}
  \begin{multline}\label{eq:Self-energy-real}
    \Re\Sigma_+(\omega)\\ = - U\left[
      \theta(\omega)\int_{-\infty}^{-\omega} +\ 
      \theta(-\omega)\int_{-\infty}^0 \right]\frac{d x}\pi
    \Im\frac{G_+(x + \omega)}{1 + \overline{U}\chi_+(x)} \\ -U
    \theta(\omega) \int_{-\omega}^0\frac{dx}\pi \Re G_+(x + \omega)
    \Im \frac 1{1 + \overline{U} \chi_+(x)}\\ -U \theta(-\omega)
    \int^{-\omega}_0\frac{dx}\pi \Im G_+(x + \omega) \Re \frac 1{1 +
      \overline{U} \chi_+(x)}\ .
  \end{multline}
  The imaginary part has a representation
 \begin{multline}\label{eq:Self-energy-imag}
   \Im\Sigma_+(\omega) = - U\left[ \theta(\omega)\int_{-\omega}^0 +\ 
     \theta(-\omega)\int^{-\omega}_0 \right]\frac{d x}\pi \\ \Im\ 
   G_+(x + \omega) \Im \frac 1{1 + \overline{U}\chi_+(x)}\ .
  \end{multline}
 \end{subequations}
 We distinguished with a subscript $+$ the way the real frequency axis
 is approached in functions with complex variables. That is,
 $G_+(x)\equiv G(x + i0^+)$, $\chi_+(x) \equiv \chi_{eh}(x + i0^+)$.
 The one-particle propagators on the r.h.s. of
 Eqs.~\eqref{eq:Self-energy} may either be bare or full propagators
 with the self-energy from the left-hand side.  Both constructions
 lead to conserving approximations when all thermodynamic quantities
 are determined from the self-energy.
 
\begin{figure}
   \includegraphics[width=7.7cm]{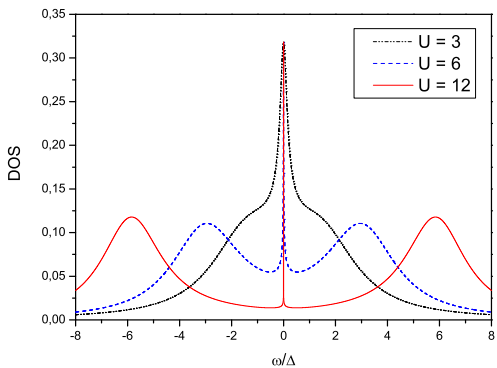}
   \caption{\label{fig:DOS-Global} (Color online) Density of states at
     zero temperature, half filling, and various interaction strengths
     without one-particle self-consistency.}
\end{figure}

\begin{figure}
   \includegraphics[width=7.7cm]{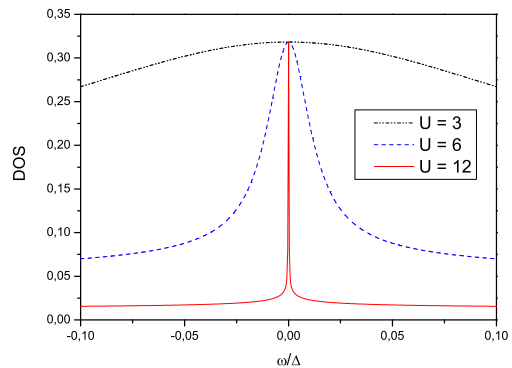}
\caption{\label{fig:DOS-local} (Color online) Detailed dependence of the
  quasiparticle peak in the density of states on the interaction
  strength.}
\end{figure}

 We first use the bare (Hartree) one-particle propagators in
 Eqs.~\eqref{eq:Self-energy}.  The density of states calculated from
 the one-particle propagator containing the self-energy from
 Eq.~\eqref{eq:Self-energy} is plotted for various interaction
 strengths in Fig.~\ref{fig:DOS-Global}.
The central Kondo quasiparticle is well formed and its width shrinks
with increasing interaction strength in accord with
Eq.~\eqref{eq:a-solution-nonself}.  See Fig.~\ref{fig:DOS-local} for a
detailed dependence of the quasiparticle peak on the interaction
strength.
Not only the central quasiparticle peak but also the satellite Hubbard
bands are well formed. Even though our approximation was justified for
small frequencies, the parquet equations with the Hartree propagators
reproduce the high-frequency features of the spectral function
surprisingly well. The Hubbard bands are positioned very close to the
exact values of the atomic levels of the impurity electrons at $\pm
U/2$.

 \subsection{One-particle self-consistency}\label{sec:1P-self-consistency}
 
\begin{figure}
   \includegraphics[width=7.5cm]{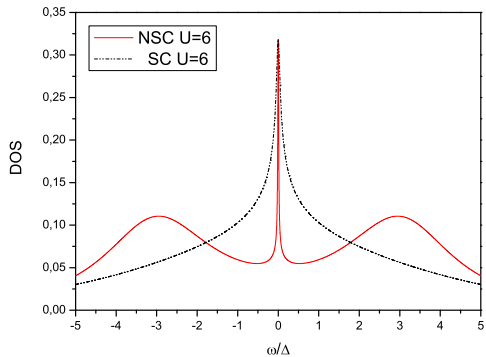}
\caption{\label{fig:DOS-Comparison}(Color online) Density of states for
  one-particle non-self-consistent (NSC) and self-consistent (SC)
  solutions compared. }
\end{figure}

 Usage of the Hartree propagators may seem inferior to a theory with
 renormalized one-particle propagators in the parquet equations. In
 fact, the opposite is true, even though the one-particle
 self-consistency adds new self-energy diagrams to those explicitly
 taken into account in the parquet construction of the self-energy
 from the two-particle vertex. The solution with renormalized
 one-particle propagators loses most of attractive features obtained
 in the parquet approach with the Hartree propagators.  First, the
 asymptotic formula for the Kondo scale,
 Eq.~\eqref{eq:a-solution-general} cannot be evaluated explicitly when
 renormalized one-particle propagators are used.  Hence, the exact
 strong-coupling asymptotics cannot be reached analytically. Second,
 the satellite peaks are completely washed out and merge with the
 central quasiparticle peak of the density of states as shown in
 Fig.~\ref{fig:DOS-Comparison}. The central peak broadens with respect
 to the solution with the Hartree propagators.  The solution with
 renormalized one-particle propagators shows overall worse agreement
 with the exact Kondo behavior derived from the Bethe ansatz. An
 analogous behavior was observed in a static simplification of the
 parquet equations.\cite{Vilk97} The failure of one-particle
 self-consistent theories to reproduce reliably the Kondo
 strong-coupling regime in impurity models is in accord with early
 analyses of SIAM using perturbation expansion in the interaction
 strength.\cite{Zlatic83}
 
 The density of states resulting from the parquet approximation with
 renormalized one-particle propagators resembles the solution of the
 renormalized RPA of Suhl.\cite{Suhl67} The central quasiparticle peak
 is, however, much broader than in the FLEX-type approximations.  It
 is even broader than that from the parquet solution with the Hartree
 propagators.  This indicates that the critical region of the
 metal-insulator transition is reached, if ever, for much stronger
 interactions in the solution with one-particle self-consistency than
 without it. The high-frequency behavior of the parquet approximation
 with renormalized one-particle propagators and the FLEX solution seem
 to behave similarly. One can prove analytically that the one-particle
 self-consistent theories with simplified one-frequency two-particle
 vertex functions behave in the high-frequency region universally and
 no Hubbard satellite bands emerge.
 
To assess the high-frequency behavior of one-particle propagators we
evaluate explicitly the leading-order contribution to the self-energy
in the critical region of the metal-insulator transition, that is, $a
\to 0$.  We obtain from Eqs.~\eqref{eq:Self-energy}
 \begin{subequations}\label{eq:SE-asymptotic}
 \begin{align}\label{eq:SE-ReS}
   \Re\Sigma_+(\omega) &\doteq \frac {U }{\overline{U} \pi^2
     \rho_0^2}\bigg[ |\ln a|\ \Re G_+(\omega) \nonumber \\ &\qquad +
   \arctan\left(\frac{\overline{U}\pi\rho_0^2\ \omega}{a}\right) \Im
   G_+(\omega)\bigg]\ , \\ \label{eq:SE-ImS} \Im\Sigma_+(\omega) &
   \doteq \frac {U}{2\overline{U} \pi^2 \rho_0^2}\ln \left[ 1 + \frac
     {\overline{U}^2 \pi^2\rho_0^4\ \omega^2} {a^2}\right] \Im
   G_+(\omega)\ .
 \end{align}
 \end{subequations}
 If we now assume $\omega \gg a\Delta$ we can further simplify the
 expression for the self-energy to
 \begin{equation}\label{eq:SE-asymptotic-large}
 \widetilde{\Sigma}_+(\omega) = \frac {U |\ln a|}{\overline{U} \pi^2
\rho_0^2} \left[ \widetilde{G}_+(\omega) + \frac {\pi\text{sign}(\omega)}{2
|\ln a|} \Im\widetilde{G}_+(\omega)\right]\ .
 \end{equation}
 We decorated the high-frequency one-particle quantities with tilde to
 distinguish them from the full ones. For tilde functions we can
 introduce a dimensionless variable
 \begin{subequations}\label{eq:1P-high}
 \begin{align}\label{eq:1P-variable}
   x &= \sqrt{\frac{\overline{U}\pi^2\rho_0^2}{U|\ln a|}}\ \omega\ .
\end{align}
We resolve the one-electron tilde propagator by using the Dyson
equation of SIAM.  We obtain an explicit solution
\begin{align}\label{eq:1P-GF}
  \widetilde{G}_+(\omega) & =
  \sqrt{\frac{\overline{U}\pi^2\rho_0^2}{U|\ln a|}}\left[\frac x2 -
    i\sqrt{1 -\frac{ x^2}4}\ \right]\ .
 \end{align}\end{subequations}
\begin{figure}
  \includegraphics[width=7.5cm]{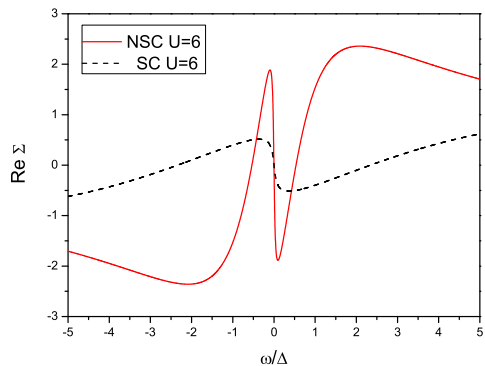}
\caption{\label{fig:ReS-Comparison}(Color online) Real part of the
  self-energy for one-particle non-self-consistent (NSC) and
  self-consistent (SC) solutions compared. }
\end{figure}
We see that one-particle self-consistent theories lead to a universal
high-frequency behavior with a semi-elliptic density of states spread
over a large interval of order $\sqrt{U|\ln
  a|/\overline{U}\pi^2\rho_0^2}$.  The FLEX solution differs only in
that $\overline{U} = U$.  The semi-elliptic form of the density of
states in high frequencies is hence universal for one-particle
self-consistent theories. The Hubbard satellite bands are completely
smeared out. The reason for the nonexistence of the satellite peaks
lies in the behavior of the real part of the self-energy in the
low-frequency region. We compared the non-self-consistent and
self-consistent solutions in Fig.~\ref{fig:ReS-Comparison}.  We can
see that the sharp low-frequency structure of the real part of the
self-energy is significantly smoothed and the peaks are broadened by
the one-particle self-consistency. Most importantly, the height of the
peaks near the Fermi energy is so much lowered that $|\omega| >
|\Re\Sigma(\omega)|$.  Only if this condition is broken, as is the
case in the non-self-consistent solution, satellite peaks emerge.

 \section{Conclusions}\label{sec:Conclusions}
 
 We presented a construction of a universal analytic impurity solver
 reliably interpolating between the weak (Fermi liquid) and strong
 (Kondo) coupling metallic regimes. The approximation is a
 simplification of two-channel parquet equations where the bounded
 irreducible vertex is replaced by a static effective interaction and
 only the low-frequency singular vertex is kept dynamical. Such a
 procedure is justified in the critical region of the metal-insulator
 transition where the low-frequency asymptotics of the singular
 two-particle vertex becomes dominant. We set with our construction
 the minimal necessary conditions that must be fulfilled to reproduce
 the Kondo exponential scale.  The Kondo behavior is observed in an
 approximate solution of SIAM if the low-frequency singularity in the
 two-particle vertex dominates the dynamics and when electron-hole and
 electron-electron multiple scatterings are self-consistently mixed in
 a balanced way.
 
 We analyzed two versions of the parquet approximation. In the first
 one, we used the Hartree one-electron propagators in the parquet
 equations for the two-particle irreducible vertices and the
 Schwinger-Dyson equation for the self-energy. In the second one we
 used the fully renormalized one-particle propagators.  Although the
 latter construction contains more Feynman diagrams renormalizing the
 self-energy, its results are less reliable than the results of the
 former approach. This conclusion is not new and is also
 understandable.\cite{Zlatic83,Vilk97} Renormalizations due to the
 one-particle self-consistency unrealistically smear and unfold the
 low-frequency structure of the self-energy.  Consequently, the
 high-frequency features of the one-particle propagator are washed out
 and no Hubbard satellite bands can be observed.
 
 One-particle self-consistency is standardly demanded in order to
 guarantee conservation laws and thermodynamic consistency. It is,
 however, not a necessary condition for approximations to be
 conserving.  Conservation laws are guaranteed in any theory with an
 approximate self-energy functional containing physical external
 sources. These sources are then used to generate the desired physical
 quantities via linear response theory. On the other hand, theories
 aiming at a description of critical phenomena such as magnetic phase
 transitions or a metal-insulator transition must be self-consistent
 in some way.  Self-consistency is necessary for any theory to handle
 singularities.  Since singularities in models with itinerant
 electrons emerge only at the two-particle level in Bethe-Salpeter
 equations, we must introduce a two-particle self-consistency. The
 parquet construction offers a very natural way to reach this goal and
 to treat singularities in the Bethe-Salpeter equations appropriately.
 One-particle self-consistency influences the critical behavior of the
 two-particle vertex, but it does not offer any direct control of
 singularities in the Bethe-Salpeter equations. It can neither
 guarantee integrability of singular vertices nor is capable to
 reproduce the correct critical behavior in the Kondo regime.  When
 used in the parquet approach, the one-particle self-consistency
 interferes in the control of the two-particle singularities from the
 Bethe-Salpeter equations achieved by the two-particle
 self-consistency.  The one-particle self-consistency can hence be
 used only when appropriately compensated so that it does not
 significantly affect the two-particle criticality. This is, however,
 not the case in the parquet approximation and the non-self-consistent
 one-particle propagators deliver better results than the
 self-consistent ones.
 
 To conclude, we derived a global approximation that is analytically
 tractable, sufficiently simple, and universal.  It can be used in a
 number of physically interesting situations, including realistic
 (multi orbital), material specific models as kind of a mean-field
 theory for models with strong local electron interaction. It may
 stand as an alternative to a recently proposed impurity solver for
 the strong-coupling regime.\cite{Dai05} Unlike the solver from
 Ref.~\onlinecite{Dai05} our approximate scheme reproduces correctly
 the Kondo strong-coupling regime.

 \section*{Acknowledgments}
 
 This research was carried out within a project AVOZ10100520 of the
 Academy of Sciences of the Czech Republic and supported in part by
 Grant No.  202/04/1055 of the Grant Agency of the Czech Republic.

\end{document}